\documentclass[12pt]{iopart}
\usepackage{graphicx}
\begin{document}
\title[Centrality dependence of elliptic flow]{The centrality
  dependence of elliptic flow at LHC}

\author{H-J Drescher$^1$, A Dumitru$^2$ 
and J-Y Ollitrault$^3$}

\address{$^1$
Frankfurt Institute for Advanced Studies (FIAS),
Johann Wolfgang Goethe-Universit\"at,
Max-von-Laue-Str.~1, 60438  Frankfurt am Main, Germany
}

\address{$^2$
Institut f\"ur Theoretische Physik,
Johann Wolfgang Goethe-Universit\"at,
Max-von-Laue-Str. 1, 60438  Frankfurt am Main, Germany
}
\address{$^3$
Service de Physique Th\'eorique, CEA/DSM/SPhT,
  CNRS/MPPU/URA2306, CEA Saclay, F-91191 Gif-sur-Yvette Cedex.}
\ead{ollitrault@cea.fr}

\begin{abstract}
We present predictions for the centrality dependence of elliptic flow 
at mid-rapidity in Pb-Pb collisions at the LHC. 
\end{abstract}
The centrality and system-size dependence of elliptic flow ($v_2$)
provides direct information on the thermalization of the matter created
in the collision. 
Ideal (non-viscous) hydrodynamics predicts that $v_2$ scales like the
eccentricity, $\varepsilon$, of the initial distribution of matter in
the transverse plane.  
Our predictions are based on this eccentricity scaling, together with
a simple parameterization of deviations from
hydrodynamics~\cite{Drescher:2007cd}: 
\begin{equation}
\label{v2k}
v_2=\frac{h\varepsilon}{1+K/0.7}, 
\end{equation}
where the scale factor $h$ is independent of system size 
and centrality, but may depend on the collision energy; 
The Knudsen number $K$ can be expressed as
\begin{equation}
\label{knud}
\frac{1}{K}=\frac{\sigma}{S}\frac{dN}{dy}\frac{1}{\sqrt{3}}.
\end{equation}
It vanishes in the hydrodynamic limit.
$dN/dy$ is the total (charged + neutral) multiplicity per unit 
rapidity, $S$ is the transverse overlap area between the two nuclei,
and  $\sigma$ is an effective (transport) partonic cross section. 

The model has two free parameters, the ``hydrodynamic limit'' $h$, and
the partonic cross section $\sigma$. The other quantities,
$\varepsilon$, $S$, $dN/dy$, must be obtained from a model for the
initial condition. Here, we choose the Color Glass Condensate (CGC)
approach, including the effect of fluctuations in the positions of
participant nucleons, which increase
$\varepsilon$~\cite{Drescher:2006ca}.  The model provides a perfect
fit to RHIC data for Au-Au and Cu-Cu collisions with $h=0.22$ and
$\sigma=5.5$~mb~\cite{Drescher:2007cd}.

We now briefly discuss the extrapolation to LHC. The hydrodynamic
limit $h$ is likely to increase from RHIC to LHC, as the QGP phase
will last longer; however, we do not have a quantitative prediction
for $h$. We predict only
the centrality dependence of $v_2$, not its absolute value. 
Figure 1 is drawn with $h=0.22$. 

The second parameter is $\sigma$, which parameterizes deviations from
ideal hydrodynamics, i.e., viscous effects. We consider two
possibilities: 1) $\sigma=5.5$~mb at LHC, as at RHIC.  2)
$\sigma\sim1/T^2$ (on dimensional grounds, assuming that no
non-perturbative scales arise), where the temperature
$T\sim(dN/dy)^{1/3}$.  This gives the value 3.3~mb in figure 1.

\begin{figure}
\centerline{\includegraphics*[width=0.7\linewidth]{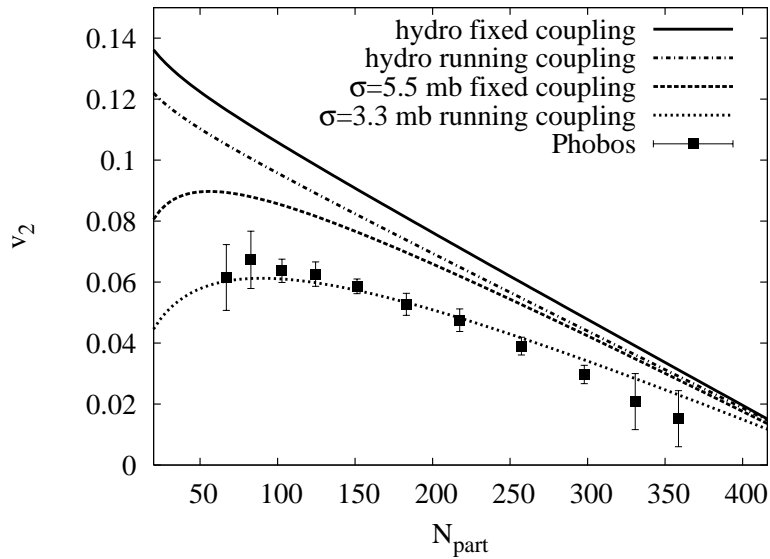}}
\caption{$v_2$ as a function of $N_{\rm part}$ at mid-rapidity for 
Pb-Pb collisions at LHC ($\sqrt{s_{NN}}=5.5$~TeV). 
\full  
and \dashddot: $\varepsilon$ scaling ($K=0$ in
(\ref{v2k})); \dashed  
and \dotted: incl.\ incomplete
thermalization, with two values of the partonic cross section.
 \fullsquare: PHOBOS data for Au-Au collisions at
RHIC~\cite{Back:2004mh}. The vertical scale is arbitrary (see text).  
\label{fig:v2lhc}}
\end{figure}
The remaining quantities ($S$, $dN/dy$ and $\varepsilon$) are obtained 
by extrapolating the CGC from RHIC to LHC, either with 
fixed-coupling (fc) or running-coupling (rc) evolution of the
saturation scale $Q_s$. 
The multiplicity per participant increases by a factor of~3 (resp.\ 2.4)
with fc (resp. rc).
The eccentricity $\varepsilon$ is 10\% larger with fc (solid curve in
figure 1) than with rc (dash-dotted curve) evolution. 
Deviations from hydrodynamics  (the $K$-dependent factor in 
Eq.~(\ref{v2k})) are somewhat smaller than at RHIC:
$v_2$ is 90\% (resp. 80\%) of the hydrodynamic limit for central
collisions if $\sigma=5.5$~mb (resp. 3.3~mb). 
Our predictions lie between the dashed and dotted curves,
up to an overall normalization factor. 
The maximum value of $v_2$ occurs for $N_{\rm part}$ between 60
($\sigma\approx$ const.) and 80 ($\sigma\sim1/T^2$).

Elliptic flow will be a first-day observable at LHC. Both 
its absolute magnitude and its centrality dependence are sensitive 
probes of initial conditions, and will help to improve our
understanding of high-density QCD.

\end{document}